# Temperature measurement of a dust particle in a RF plasma GEC reference cell

Jie Kong, Ke Qiao, Lorin S. Matthews and Truell W. Hyde

*Center for Astrophysics, Space Physics, and Engineering Research (CASPER)*

*Baylor University, Waco, Texas 76798-7310, USA*

Abstract

The thermal motion of a dust particle levitated in a plasma chamber is similar to that described by Brownian motion in many ways. The primary differences between a dust particle in a plasma system and a free Brownian particle is that in addition to the random collisions between the dust particle and the neutral gas atoms, there are electric field fluctuations, dust charge fluctuations, and correlated motions from the unwanted continuous signals originating within the plasma system itself. This last contribution does not include random motion and is therefore separable from the random motion in a 'normal' temperature measurement. In this paper, we discuss how to separate random and coherent motion of a dust particle confined in a glass box in a Gaseous Electronic Conference radio frequency reference cell employing experimentally determined dust particle fluctuation data analyzed using the mean square displacement technique.

1. Introduction

The coupling parameter $\Gamma$ for a dusty plasma system is defined as the ratio of the interparticle potential energy to the dust kinetic (thermal) energy [1 – 3]. A two dimensional dust system exhibits a phase transition from a liquid to crystalline state as the coupling parameter increases beyond a critical value, $\Gamma > \Gamma_c$, where $\Gamma_c$ is approximately 170 [4 – 6]. To determine this system coupling parameter experimentally, a proper measurement of the dust kinetic energy, i.e., the dust temperature, is very important. By definition, the temperature of a dust particle is taken to be

$$k_B T = m \langle v^2 \rangle \tag{1}$$

where $k_B$ is the Boltzmann constant, $m$ is the dust mass, and $\langle v^2 \rangle$ is the mean square velocity of the random motion of the dust particle. Therefore, an accurate determination of $\langle v^2 \rangle$ is crucial for measurement of the dust temperature. There are different techniques to determine $\langle v^2 \rangle$, such as using the velocity distribution (which is often assumed to be a Gaussian distribution under normal dusty plasma conditions), where $\langle v^2 \rangle$ represented the standard deviation, and using the autocorrelation function (ACF) and assumed ballistic motion at short time scales of the mean square displacement (MSD) [7 – 10]. The experimentally measured dust temperatures determined employing either of these techniques are much higher than that of the neutral gas [11 – 17]. Possible explanations for this include charge fluctuations [18 – 22], and various plasma-dust instabilities in the electric field of the gas discharge chambers [23, 24].

In addition to the random motion created by collisions between a dust particle and neutral gas molecules, and stochastic electrostatic and charge fluctuations, dust particles are also perturbed by oscillations imposed due to continuous driving sources. Quinn and Goree pointed out in [18] that the measured mean square velocity $\langle v^2 \rangle$ includes two main components due to random motion and coherent motion, where the latter is caused by correlated waves created within the plasma system. Unfortunately, how to separate these two parts remains an unanswered question. In this paper, we will explain how to obtain a measurement of the dust particle kinetic temperature from only the random motion using the MSD technique. A brief theoretical background for this technique will be given in Section 2. Experimental results and discussions are presented in Sections 3 and 4 respectively, with conclusions in Section 5.

2. Theoretical background

It has long been known that Brownian particle can be used as a probe to determine the properties of its environment. In one of his seminal papers, Einstein related the mean square displacement (MSD) of a free Brownian particle over a time $\Delta t$ to the diffusion constant $D$ as [25]

$$\langle x^2 \rangle = 2D\Delta t \tag{2}$$

where $D = \mu k_B T$, with $\mu$ defined as the mobility. This relationship is only valid for time intervals $\Delta t \gg \tau_p$, where $\tau_p$ is the momentum relaxation time. At very short time scales ($\Delta t \ll \tau_p$) particle motion may be considered to be ballistic, as given by

$$\langle x^2 \rangle = \langle v^2 \rangle \Delta t^2 = (k_B T / m) \Delta t^2 \tag{3}$$

which characterizes the short time scale MSD for a Brownian particle.

Eqs 2 and 3 are derived assuming non-bounded particles, i.e., the Langevin equation for describing the particle motion is free of any confinement force [26, 27]

$$m\dot{v} = -m\gamma v + R(t) \tag{4}$$

where $R(t)$ is the fluctuating force and $\gamma = 1/\tau_p$ is the damping coefficient [28].

For dust particles confined in a harmonic potential well, Eq 4 is modified to read as [29]

$$m\ddot{x} = -m\gamma \dot{x} - kx + R(t) \tag{5}$$

where $k = m\omega_0^2$ and $\omega_0$ is the particle resonance frequency. The MSD solution of Eq 5 is [10, 30] (also see the Appendix A1 – A8),

$$\langle x^2 \rangle = A_0 \left[ 1 - \exp\left(-\frac{\gamma \Delta t}{2}\right) \left\{ \cos(\hat{\omega}\Delta t) - \frac{\gamma}{2\hat{\omega}} \sin(\hat{\omega}\Delta t) \right\} \right] \tag{6}$$

where $A_0 = \dfrac{2k_B T}{m\omega_0^2}$ and $\hat{\omega} = \sqrt{\omega_0^2 - \left(\dfrac{\gamma}{2}\right)^2}$.

Eq 6 clearly shows that as $\Delta t$ increases to $\Delta t \gg 1/\gamma$ [31, 32],

$$\langle x^2 \rangle_{\Delta t \gg 1/\gamma} = A_0 = \frac{2k_B T}{m\omega_0^2} \tag{7}$$

As can be seen, instead of being linearly proportional to $\Delta t$ as in Eq 1, $\langle x^2 \rangle$ is now a constant which is related to both the kinetic temperature and the resonance frequency of the particle and is independent of $\Delta t$. Experimentally the constant $A_0$ is very easy to extract as will be shown in the following section.

However, Eq 5 is based on an ideal system employing a harmonic confinement. For a dusty plasma system with unwanted continuous oscillations, Eq 5 must be modified as,

$$m\ddot{x} + m\gamma\dot{x} + kx+ = R(t) + \sum_{Correlated} a_i \cos \omega_i t \tag{8}$$

where $a_i$ and $\omega_i$ are the amplitude and frequency of individual oscillations within the system. These unwanted oscillations may be mechanical or electronic. The corresponding solution for Eq 8 is (see A10 and A11 in the Appendix),

$$\langle x^2 \rangle = A_0 \left[ 1 - \exp\left(-\frac{\gamma \Delta t}{2}\right) \left\{ \cos(\hat{\omega}\Delta t) - \frac{\gamma}{2\hat{\omega}} \sin(\hat{\omega}\Delta t) \right\} \right] + \sum_{Correlated} C_i \cos(\omega_i \Delta t + \varphi_i) \tag{9}$$

Eq 9 indicates that when $C_i \ll A_0 = \frac{2k_B T}{m\omega_0^2}$, and $\Delta t \gg 1/\gamma$, the mean square displacement approaches an equilibrium value $A_0$ with small modulations about this value of frequency $\omega_i$. This means that these oscillations will not affect the constant $A_0$, which is related to the dust temperature. The implication of this is that the experimentally determined average MSD at $\Delta t \gg 1/\gamma$ is not affected by this continuous oscillation. Therefore, by measuring the constant $A_0$ the stochastic fluctuation can be separated from the correlated oscillations.

The kinetic energy supplied by the continuous oscillations to the dust particle is

$$E_{Corr} = \sum_{Corr} \frac{1}{2} m\omega_i^2 \Delta_i^2 \tag{10}$$

where $\Delta_i$ is the i$^{th}$ oscillation amplitude. Because this kinetic energy is proportional to the square of the oscillation frequency, a greater contribution comes from higher frequency oscillations when the amplitudes of all oscillations are similar.

The following sections describe a recent experiment which uses the natural random motion of a single dust particle confined within a glass box placed on the lower powered electrode in a Gaseous Electronic Conference (GEC) rf reference cell to verify Eqs 8 and 9. This is accomplished by measuring the particle's mean square displacement and then using this data to derive both the oscillation frequency (i.e., the confinement force constant) and the temperature of the dust particle.

3. Experiment and results

The experiments described here were conducted in one of CASPER's GEC rf reference cells [33]. Melamine formaldehyde (MF) dust particles having a diameter of 8.89 µm were introduced into a glass box of dimension 10.5 mm × 10.5 mm × 12.5 mm (width × length × height) placed on the lower powered electrode using a dust shaker mounted above the upper ring electrode. The number of dust particles confined within the box was controlled by adjusting the system's rf power. A single confined dust particle was used for this experiment. For all experiments, a side mounted high speed camera recorded 60 seconds of dust particle motion at 500 frames per second (fps) (illuminated using a 50 mW solid state laser at 660 nm), neutral gas pressure was held at 13.3 Pa and rf power was 2.25 W. An important aspect of the experimental setup is that the DC bias of the lower electrode can be modulated using a function generator. This allows a signal, consisting of a single frequency or random noise, to be sent to the lower electrode in order to generate either *correlated* or *random* dust particle motion. For a single frequency input, the frequency selected should be far from the dust particle's intrinsic frequency, $\omega_0$, in order to avoid resonance and reduce the mode coupling effects. In this experiment, a single frequency input of 110 Hz was chosen. Adjusting the driving voltage allows the amplitude of the single frequency or random noise to be controlled. Therefore, the values of $C_i = C_{110Hz}$, and $A_0$ in Eq 9 can be varied independently, i.e., $A_0$ and $C_{110Hz}$ are now independent functions of the driving voltage $V_{drive}$. The original raw data (i.e., photos) are processed using "ImageJ" developed at the National Institutes of Health [34]. Fig 1 shows representative raw data of dust particle position fluctuations. The random noise driving amplitude was 2000 mV to show the difference between the amplitude of induced vertical and horizontal oscillations.

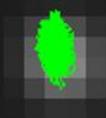

(a)

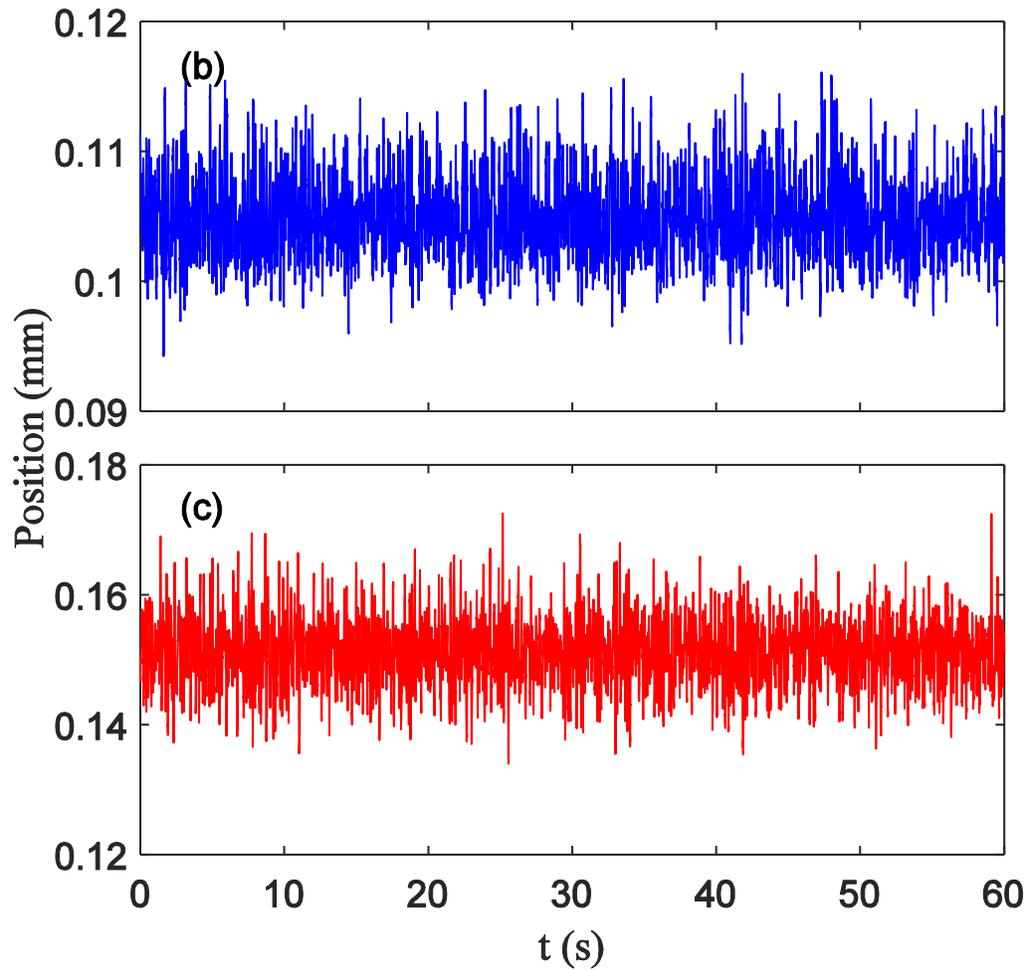

Fig 1. Experimental data for a single dust particle's position fluctuations. (a) Raw photo with detected particle trajectory superimposed in green. (b) Horizontal and (c) vertical fluctuations as a function of time derived from (a).

The particle's mean square displacement (MSD) can be calculated from the experimental data shown in Fig 1, and the dust particle's corresponding temperature $T$, resonance frequency $\omega_0$ and damping coefficient $\gamma$ then derived using the theoretical fit provided by Eq 6. An example MSD (under the conditions of no applied DC bias perturbation) is shown in Fig 2.

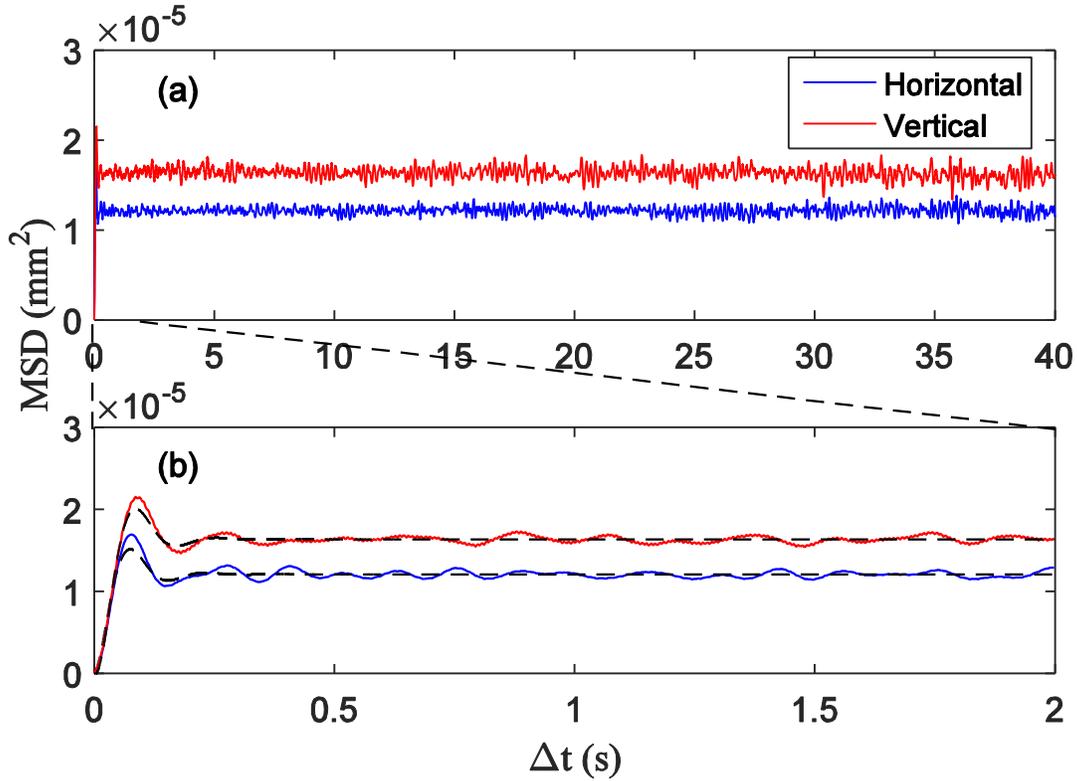

Fig 2. (a) Overview of a representative experimental MSD data set. (b) Expanded view of (a) for $\Delta t \leq 1.0$ s. Solid lines are experimental values and the dashed lines are the theoretical fit calculated using Eq 6.

As can be seen in Fig 2, the MSDs are flat for a region $0.5 \leq \Delta t \leq 40$ s. This constant value is $A_0$, which can be obtained by averaging over at least $2 \times 10^4$ data points under the experimental setup of camera rate at 500 fps for 60 sec.

Fig 3 shows Fast Fourier Transformation (FFT) spectra for dust particle positions having different values of $V_{drive}$ for both 110 Hz single frequency and random noise DC bias modulations. The single frequency driving peak-to-peak voltage is measured before a 20 dB attenuator.

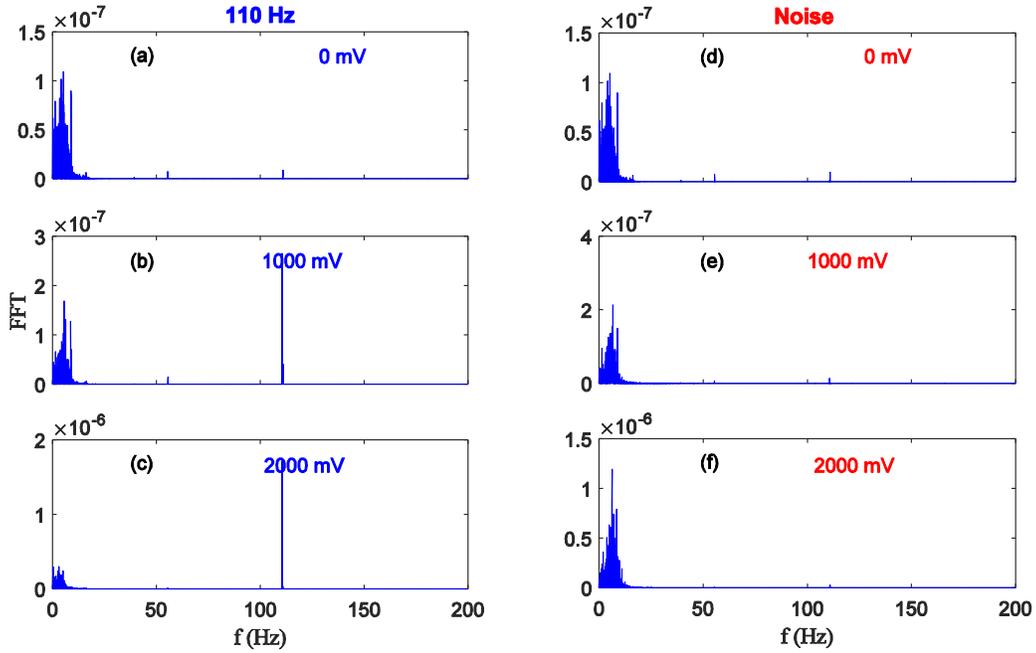

Fig 3. FFT spectra of particle motion produced by modulation of the DC bias of the lower electrode using a single frequency (a) – (c) and random noise (d) – (f). The amplitude of the modulation is controlled by the driving voltage, as indicated in each panel. Only results for the vertical direction are shown as there are no significant changes to the spectra for motion in the horizontal direction in either case.

As can be seen in Fig 3, there is only minimal increase within the low frequency band (< 10 Hz, where the dust intrinsic frequency, $\omega_0$, is located) as the single frequency driving voltage increases (notice the difference in vertical scaling for each panel), while increasing the random noise driving voltage has a strong effect on the low frequency band.

The effect of the modulation of the DC bias on the MSDs in the horizontal and vertical motion is illustrated in Figure 4. The modulation of the DC bias using a single 110 Hz frequency had very little effect on either the horizontal (4a) or vertical (4b) motion, and was only weakly dependent on the magnitude of the driving amplitude. However, modulation of the DC bias employing random noise increased the MSD in the vertical direction, with the magnitude of this increase proportional to the driving amplitude (4d). There was no correlated effect on the MSD in the horizontal direction (4c).

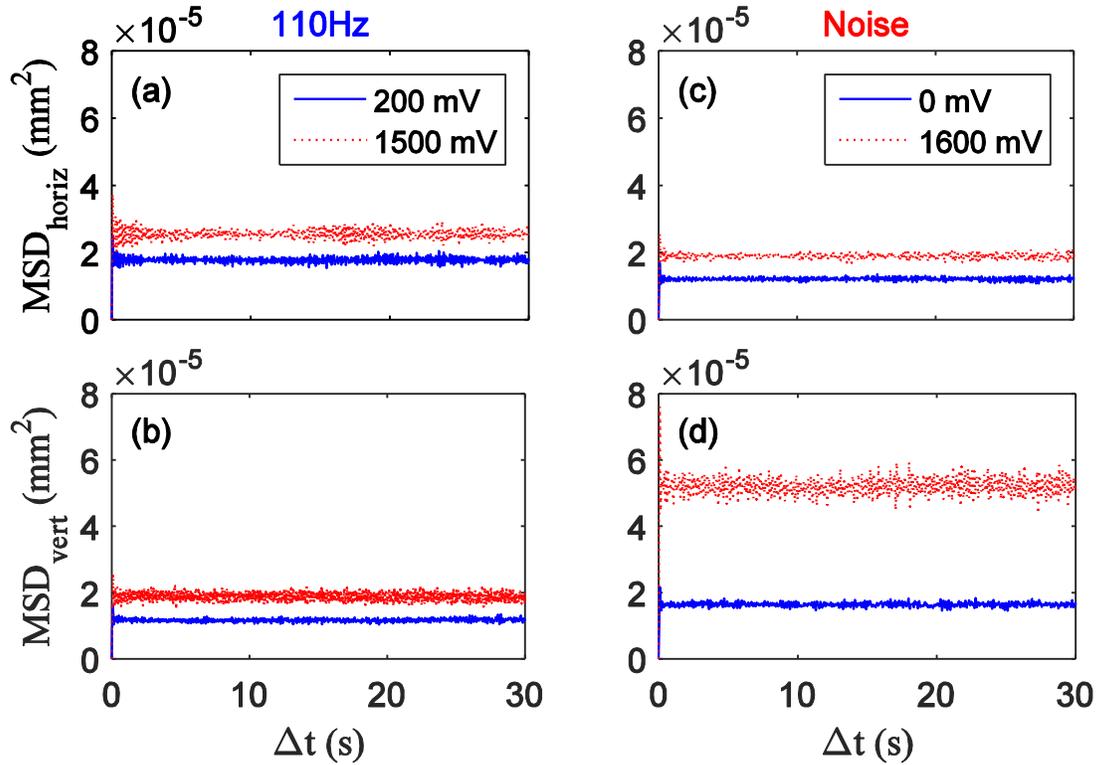

Fig 4. MSD for a 110 Hz single wavelength driving voltage at 200 mV and 1500 mV in the (a) horizontal and (b) vertical directions, respectively. MSD for a random noise driving voltage at 0 mV and 1600 mV in (c) horizontal and (d) vertical directions, respectively.

As can be seen in Fig 4, horizontal (a) and vertical (b) equilibrium MSDs are only minimally affected by changing the amplitude of a single-frequency modulation. On the other hand, changing the amplitude of the random noise modulation causes the vertical equilibrium position to shift significantly (d), while leaving the horizontal equilibrium MSD relatively unchanged (c).

Examining the MSD at short time scales reveals the small amplitude oscillations imposed in the vertical direction by the 2000 mV, 110 Hz DC bias modulation (Fig 5), while the MSD in the horizontal direction remains unaffected by the single frequency modulation.

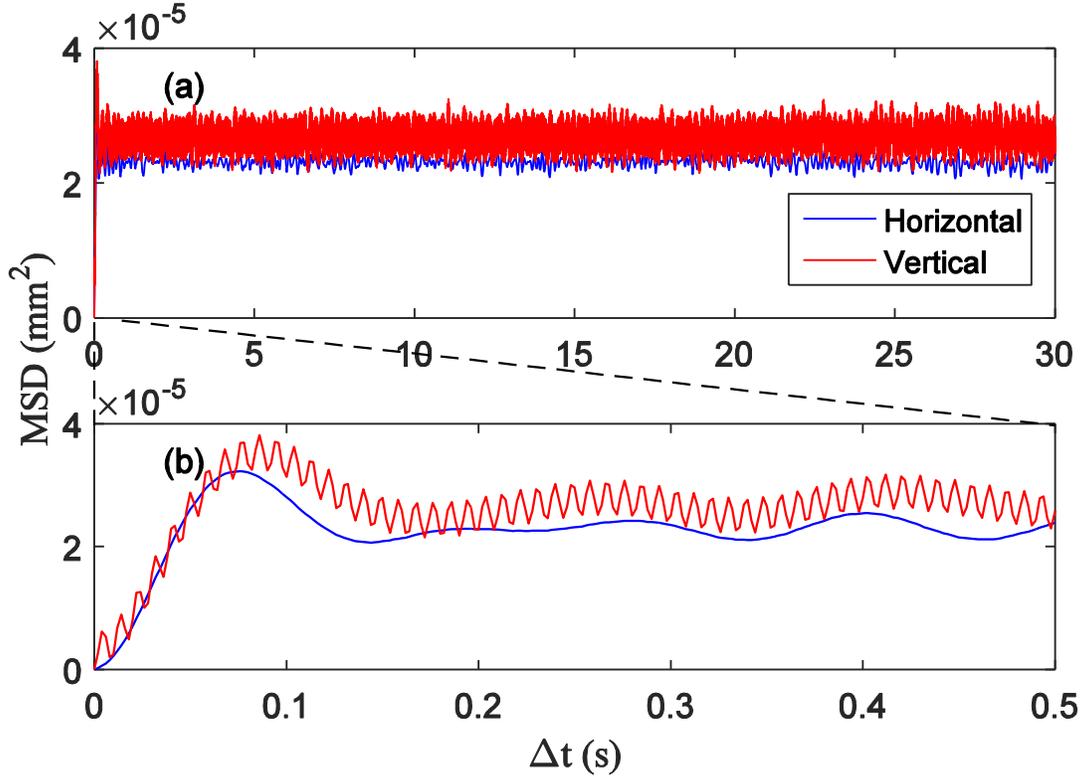

Fig 5. (a) Comparison of the horizontal and vertical MSD for a 110 Hz single wavelength modulation of the DC bias provided to the lower electrode. (b) Expanded view showing short time scales.

As can be seen in Fig 5, the constant $A_0$ can still be obtained by averaging the MSD over $\Delta t \geq 0.5$ s. However, the short time scale MSD is strongly affected by the 110 Hz oscillation in the vertical direction.

4. Discussion

To verify Eq 9 experimentally, we must first prove that a random driving force is related to the equilibrium value of the MSD. Since $A_0 = \dfrac{2k_B T}{m\omega_0^2}$, and thus is directly related to the dust kinetic temperature, a positive correlation between $A_0$ and the amplitude of the random driving force would imply that the random driving force also contributes to the dust kinetic temperature.

The dust particle temperature is derived using Eq 6. For this case, the drag will be assumed to be to be given by the Epstein drag [28] in order to simplify the fitting process, with the drag coefficient given by

$$\gamma = \delta \frac{8p}{\pi r_d \rho_d v_{th}} \qquad (11)$$

where the coefficient for diffuse reflection is $\delta = 1.44$ for MF dust particles in argon gas [35], $r_d$ and $\rho_d$ are the dust particle radius and density respectively, $v_{th}$ is the thermal velocity of the neutral gas, and $p$ is the pressure.

The equilibrium value $A_0$ and the resonance frequency $\omega_0$ can now be determined using Eq 6 to fit the experimentally derived MSDs shown in Figures 2 and 5. The resonance frequencies found in this manner are $f_{0horiz} = \omega_{0horiz}/2\pi = 7.0$ Hz, $f_{0vert} = 6.6$ Hz (for the non-driven oscillations shown in Fig 2), and $f_{0horiz} = \omega_{0horiz}/2\pi = 7.8$ Hz, $f_{0vert} = \omega_{0horiz}/2\pi = 6.5$ Hz (for the single frequency driven oscillations shown in Fig 5). The dust temperatures derived as a function of the driving amplitude are shown in Fig 6 for both the random and single frequency driving signals.

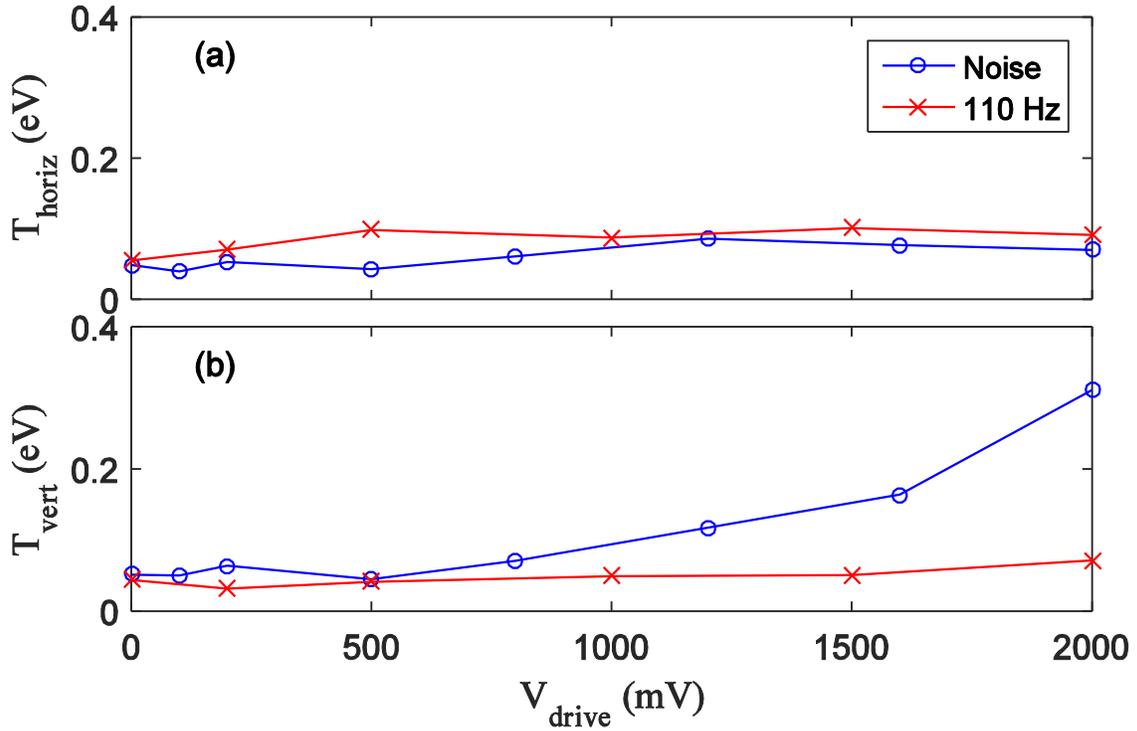

Fig 6. The calculated dust temperature as a function of the driving amplitude for both a single frequency and random driving signal in the (a) horizontal, and (b) vertical direction. Connecting lines serve to guild the eye.

It can be seen that the dust temperature in the vertical direction increases as the driving amplitude of the noise increases, while remaining almost constant in the horizontal direction. This is due to the fact that DC modulation of the lower powered electrode creates a variation in the confining electric field primarily in the vertical direction. In the vertical direction, the supporting electric field force is balanced by the gravitational force acting on the dust particle, which is constant. Therefore, changes in the vertical electric field represent an asymmetric driving force which changes the instantaneous vertical equilibrium position of the particle, which is added to the natural fluctuation about the equilibrium position. Changing the DC bias of the lower electrode contributes a much smaller variation to the horizontal confining fields, with the change being symmetric to each side. Thus the horizontal equilibrium position is not changed.

Secondly, it must be proved that a driving force consisting of a continuous single frequency wave of constant amplitude, represented by $\omega_i$ and $C_i$ in Eq 9, does not contribute to the dust temperature. In this case, the kinetic temperature of the dust particle should not change as the input amplitude of the continuous wave increases (as long as $C_i \ll A_0$), since the continuous wave should only induce small oscillations around the equilibrium value $A_0$. As shown in Fig 5, a single frequency driving force imposes a modulation on the MSD in the vertical direction with the same frequency as the driving force. This modulation does not significantly change the equilibrium value $A_0$, but it does make calculation of the ballistic motion over the short time regime ($\Delta t \ll 1/\gamma = 0.028\,\text{s}$ for this experiment) very difficult if not impossible. Therefore, employing the ballistic motion assumption, Eq 3, to calculate the dust temperature can be easily affected by any unwanted coherent motion from the plasma system. On the other hand, as pointed out in the previous section, the constant $A_0$ can be calculated by averaging over a large percentage of the collected data. This is another big advantage over the ballistic motion method, which only consists a few data points.

In addition to the MSD and the short time scale techniques, the dust particle kinetic energy can be obtained using the velocity probability distribution function (PDF). Representative velocity PDFs from this experiment are shown in Fig 7. As shown, Gaussian distributions fit the experimental data well in both the horizontal and vertical directions.

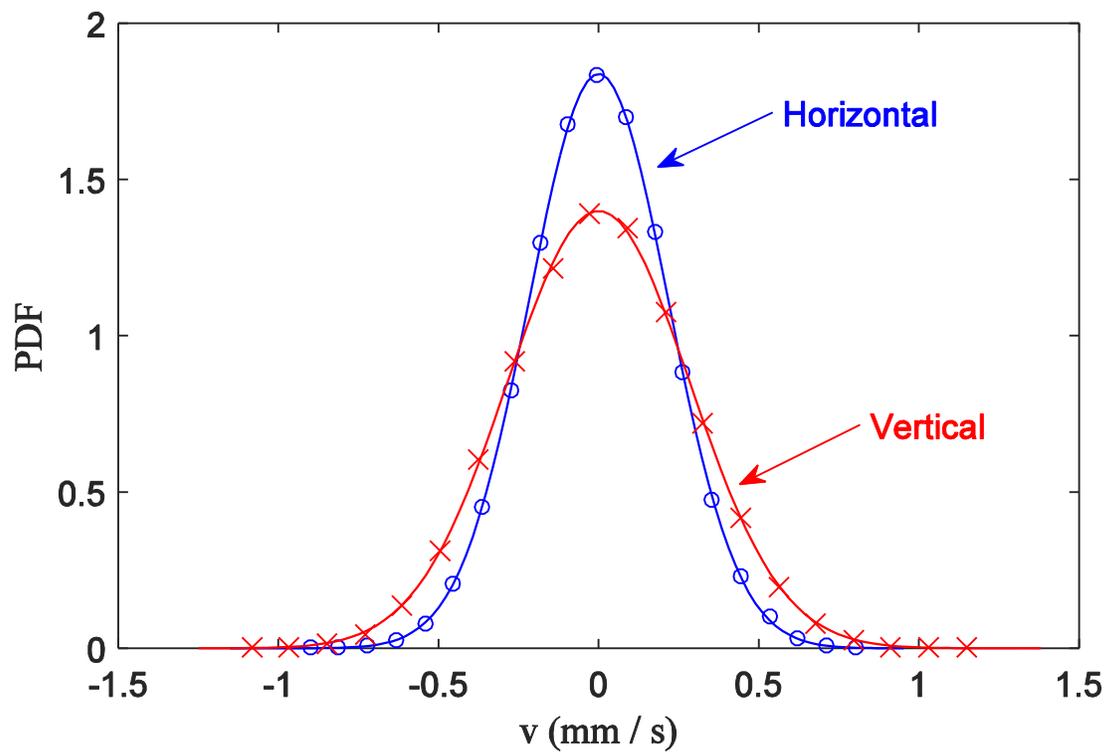

Fig 7. Representative velocity probability distribution functions (normalized) for driven noise modulation with a driving amplitude of 100 mV. Symbols represent experimental values while solid lines provide a theoretical Gaussian distribution fit.

The temperatures calculated from the Gaussian fit, where the standard deviation gives a measure of $\langle v^2 \rangle_{Gauss}$ which is related to the temperature by Eq 1, is shown in Fig 8 as a function of the driving amplitude. This is compared to the temperatures calculated by the MSD technique.

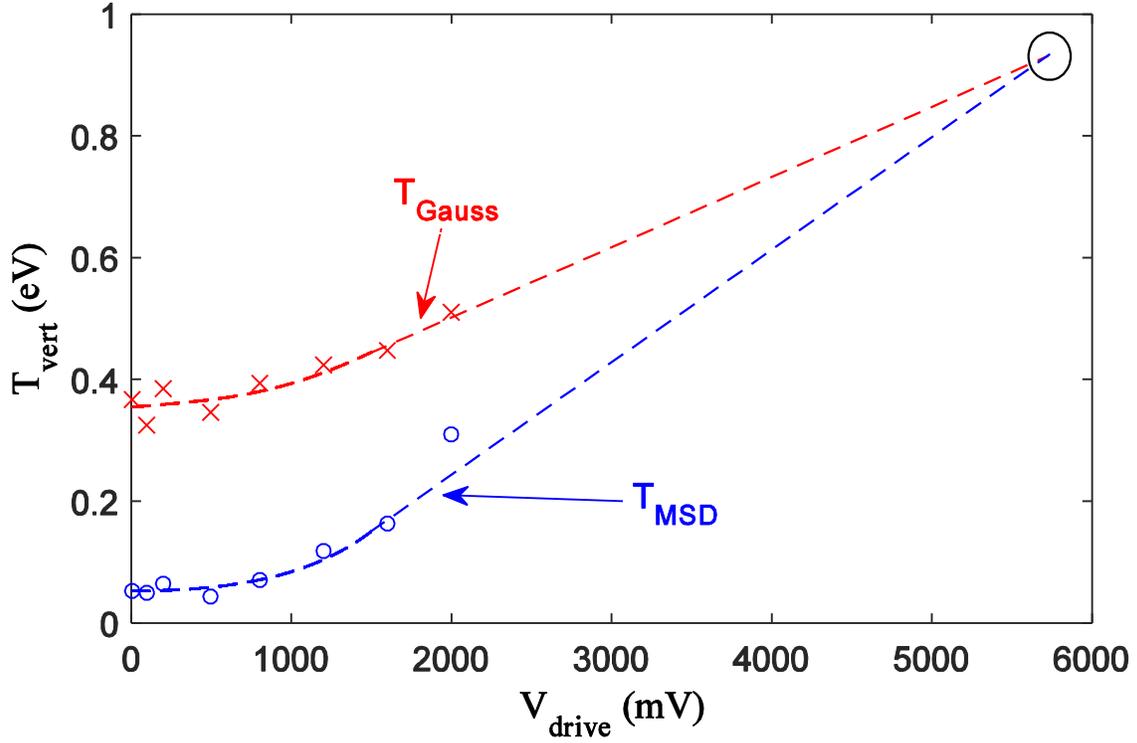

Fig 8. Comparison of dust temperature derived using MSD technique (circles) and velocity distribution function technique (crosses). Oscillations are driven using random noise.

The values for the temperature at a driving amplitude of 0 mV in Fig 8a are 0.052 eV and 0.35 eV as derived from the MSD and PDF of velocities respectively, which correspond to approximately 600 K and 4000 K. Extrapolating the data to larger driving amplitudes as shown (see the circled area around 5700 mV in Fig 8), the two extrapolated temperatures converge at a certain point. This can be explained by assuming

$$T_{Gauss} = T_{real} + T_{correlated} \qquad (12)$$

where $T_{real} = T_{MSD}$ and $T_{correlated}$ represents all the correlated oscillation contributions. Noting that increasing the noise amplitude only increases $T_{MSD}$, as the noise level increases to a point where $T_{correlated}$ can be ignored, eventually $T_{Gauss} = T_{MSD}$.

As mentioned earlier, the continuous driving sources add kinetic energy to the dust particle (Eq 10). If not separated from the stochastic fluctuations, these oscillations lead to in apparent increase in the dust temperature. The energy contributed by each oscillation frequency is represented by the amplitude of the FFT spectra. It can be seen in Figure 3 that the system (without added single-frequency or noise driven sources) shows small peaks at 55 Hz and 110 Hz. The amplitude of these oscillations is about 1/10 of the low frequency band ($<10$ Hz). Assuming that the contribution due to each of these frequencies is $\Delta h/10$, where $\Delta h$ is the standard deviation of the displacement for the non-driven case (Fig 3a). The total contribution to the temperature can be calculated using Eqs 10 and 1. The excess temperature associated with the 55 Hz is $T_{55} = 560$ K, and the excess temperature associated with the driving frequency 110 Hz is $T_{110} = 2230$ K. Adding these to the temperature derived from $A_0$ using the MSD method, the total temperature is $T_{total} = 3400$ K, close to the temperature determined from the Gaussian fit to the velocity distribution shown in Figure 8, $T_{Gauss} = 4000$ K.

5. Conclusion

In this paper, temperature measurement of a dust particle in a dusty plasma chamber is discussed in detail. Based on a MSD analysis, the contribution to the temperature measurement from random fluctuations of a dust particle confined in a glass box in a GEC rf reference cell is separated from the motion of a continuous single frequency perturbation. Theoretical analysis and experimental data show that the equilibrium MSD at $\Delta t >> 1/\gamma$ is a function of the amplitude of the driving random noise, but independent of the amplitude of a continuous single frequency perturbation. Thus, a temperature derived using this method will be lower than that using the velocity PDF method, where both the random and correlated motions are included in the particle velocities, and thus a measurement of temperature is based on the energy of both random and correlated motions. A real system will have particle motion driven by both random forces (Brownian motion, fluctuations of the particle charge, fluctuations in the electric field, etc.) as well as motion which is correlated with driving forces at a single frequency. It is important to note that the MSD technique yields a temperature which includes energy contributions from all random effects. However, it is not yet known how to distinguish between the contributions of each of these effects which together form what is commonly referred to as the temperature of the dust particle.

References

1. E. Wigner, "Effects of the electron interaction on the energy levels of electrons", Trans. Faraday Soc. 34, 678, (1938).

2. H. Thomas, G. E. Morfill, V. Demmel, and J. Goree, "Plasma crystal: Coulomb crystallization in a dusty plasma", Phys. Rev. Lett., 73, 652 – 655, (1994).

3. S. Ichimaru, "Strongly coupled plasmas: high-density classical plasmas and degenerate electron liquids", Rev. Mod. Phys. 54, 1017, (1982).

4. R. T. Farouki and S. Hamaguchi, "Thermodynamics of strongly-coupled Yukawa systems near the one-component-plasma limit. II. Molecular dynamics simulations", J. Chem. Phys., 101, 9885, (1995).

5. Frank Melandso, "Heating and phase transitions of dust-plasma crystals in a flowing plasma", Phys. Rev. E 55, 7495, (1997).

6. Niels Otani and A. Bhattacharjee, "Debye Shielding and Particle Correlations in Strongly Coupled Dusty Plasmas", Phys. Rev. Lett., 78, 1468, (1997).

7. Tongcang Li, Simon Kheifets, David Medellin, Mark G. Raizen, "Measurement of the instantaneous velocity of a Brownian particle", Science, 328, 1673 – 1675, (2010).

8. Simon Kheifets, Akarsh Simha, Kevin Melin, Tongcang Li, Mark G. Raizen, "Observation of Brownian motion in liquids at short times: instantaneous velocity and memory loss", Science, 343, 1493 – 1496, (2014).

9. Peter N. Pusey, "Brownian motion goes ballistic", Science, 332, 802 – 803, (2011).

10. Christian Schmidt and Alexander Piel, "Stochastic heating of a single Brownian particle by charge fluctuations in a radio-frequency produced plasma sheath", Phys. Rev. E 92, 043106, (2015).

11. Hubertus M. Thomas and Gregor E. Morfill, "Melting dynamics of a plasma crystal", Nature, 379, 806, (1996).

12. A. Melzer, A. Homann, and A. Piel, "Experimental investigation of the melting transition of the plasma crystal", Phys. Rev. E 53, 2757, (1996).

13. Jeremiah D. Williams and Edward Thomas, Jr., "Initial measurement of the kinetic dust temperature of a weakly coupled dusty plasma", Phys. Plasmas, 13, 063509, (2006).

14. R. A. Quinn and J. Goree, "Single-particle Langevin model of particle temperature in dusty plasmas", Phys. Rev. E 61, 3033-3041, (2000).

15. Amit K. Mukhopadhyay and J. Goree, "Two-Particle distribution and correlation function for a 1D dusty plasma experiment", Phys. Rev. Lett., 109, 165003, (2012).

16. Amit K. Mukhopadhyay and J. Goree, Phys. Rev. Lett., 111, 139902, (2013).

17. J. B. Pieper and J. Goree, "Dispersion of plasma dust acoustic waves in the strong-coupling regime", Phys. Rev. Lett., 77, 3137, (1996).

18. R. A. Quinn and J. Goree, "Experimental investigation of particle heating in a strongly coupled dusty plasma", Phys. Plasmas, 7, 3904-3911, (2000).

19. V. V. Zhakhovski, V. I. Molotkov, A. P. Nefedov, V. M. Torchinski, A. G. Khrapak, and V. E. Fortov, "Anomalous heating of a system of dust particles in a gas-discharge plasma", JETP Lett., 66, 419 – 425, (1997).

20. O. S. Vaulina, S. V. Vladimirov, A. Yu. Repin, and J. Goree, "Effect of electrostatic plasma oscillations on the kinetic energy of a charged macroparticle", Phys. Plasmas, 13 012111, (2006).

21. O. S. Vaulina, S. A. Khrapak, A. P. Nefedov, and O. F. Petrov, "Charge-fluctuation-induced heating of dust particles in a plasma", Phys. Rev. E 60, 5959 – 5964, (1999).

22. G. E. Morfill and H. Thomas, "Plasma crystal", J. Vac. Sci. Technol. A 14, 490, (1996).

23. O. S. Vaulina, Plasma Phys. Rep., "Transport properties of nonideal systems with isotropic pair interactions between particles", 30, 652-661, (2004).

24. O. S. Vaulina, A. A. Samarian, B. James, O. F. Petrov, and V. E. Fortov, "Analysis of macroparticle charging in the near-electrode layer of a high-frequency capacitive discharge", J. Experimental and Theoretical Physics, 96, 1037 – 1044, (2003).

25. Albert Einstein, "Uber die von der molekularkinetischen Theorie der Warme geforderte Bewegung von in ruhenden Flussigkeiten suspendierten Teilchen", Ann. Phys. (Berlin) 322, 549–560 (1905).

26. Ryogo Kubo, "The fluctuation – dissipation theorem", Rep. Prog. Phys., 29, 255 – 283, (1966).

27. Ryogo Kubo, "Brownian motion and nonequilibrium statistical mechanics", Science, 233, 330 – 334, (1986).

28. Paul S. Epstein, "On the resistance experienced by spheres in their motion through gases", Phys. Rev. 23, 710 – 733, (1924).

29. M. C. Wang and G. E. Uhlenbeck, "On the theory of the Brownian motion II", Rev. Mod. Phys., 323 – 342, (1945).

30. Tongcang Li and Mark G. Raizen, "Brownian motion at short time scales", Ann. Phys. (Berlin), 525, 281 – 295, (2013).

31. V. Nosenko and J. Goree, "Laser method of heating monolayer dusty plasmas", Phys. Plasmas, 13, 032106, (2006).

32. Jie Kong, Ke Qiao, Lorin Matthews, and Truell W. Hyde, "Interaction force in a vertical dust chain inside a glass box", Phys. Rev. E 90, 013107, (2014).

33. Truell W. Hyde, Jie Kong, and Lorin S. Matthews, "Helical structures in vertically aligned dust particle chains in a complex plasma", Phys. Rev. E 87, 053106 (2013).

34. Wayne Rasband, National Institutes of Health, USA (http://imagej.nih.gov/ij).

35. Hendrik Jung, Franko Greiner, Oguz Han Asnaz, Jan Carstensen, and Alexander Piel, "Exploring the wake of a dust particle by a continuously approaching test grain", Phys. Plasmas, 22, 053702, (2015).

36. Gregory H. Wannier, Statistical Physics, John Wiley and Sons, New York, 1966.

37. Paul Langevin. Sur la théorie du mouvement brownien. C. Rendus Acad. Sci. Paris, 146:530–533, 1908. Translated version: Don S. Lemons and Anthony Gythiel, Am. J. Phys., 65, 1079 – 1081, (1997).

38. Robert Zwanzig. "Nonequilibrium statistical mechanics", Oxford University Press, 2001.

39. Gerald R. Kneller, "Anomalous diffusion in biomolecular systems from the perspective of non-equilibrium statistical physics", Act Phys. Pol. B46, 1167 – 1199, (2015).

40. Gerald R. Kneller, "Stochastic dynamics and relaxation in molecular systems – Brownian dynamics and beyond", http://dirac.cnrs-orleans.fr/~kneller/SOCRATES/lecture.pdf.

Appendix

For a particle confined by a harmonic potential well, the Langevin equation is [36 – 39],

$$m\dot{v} = -m\gamma v - m\omega_0^2 x + R(t) \tag{A1}$$

where $\gamma$ is the damping coefficient, $\omega_0$ is the dust resonance frequency, $R(t)$ is the random force, and $m$ is the dust particle mass. Divide both sides by $m$,

$$\dot{v} + \gamma v + \omega_0^2 x = +r(t) \tag{A2}$$

where $r(t) = R(t)/m$. A2 can be rewritten as,

$$\dot{v} + \gamma v + \omega_0^2 \int_0^t v(\tau) d\tau = +r(t) \tag{A3}$$

Multiplication with $v(0)$ and averaging over time yields [40]

$$\dot{c}_{vv} + \gamma c_{vv} + \omega_0^2 \int_0^t c_{vv}(\tau) d\tau = 0 \tag{A4}$$

where $c_{vv} = \langle v(t)v(0) \rangle_\tau$ is the velocity autocorrelation function (VACF), and $\langle v(0)r(t) \rangle_\tau = 0$. Applying Laplace transform to A4, the VACF is solved as,

$$\hat{c}_{vv}(s) = \frac{k_B T}{m} \frac{s}{(s-s_1)(s-s_2)} \tag{A5}$$

where $\hat{c}_{vv}(s)$ is the Laplace transform of VACF, $s_{1,2} = -\frac{\gamma}{2} \pm i\hat{\omega}$, $\hat{\omega} = \sqrt{\omega_0^2 - \left(\frac{\gamma}{2}\right)^2}$. The inverse Laplace transform of A5 is the VACF

$$c_{vv}(t) = \frac{k_B T}{m} \exp\left(-\frac{\gamma}{2}t\right) \left\{ \cos(\hat{\omega}t) - \frac{\gamma}{2\hat{\omega}} \sin(\hat{\omega}t) \right\} \tag{A6}$$

To derive the MSD solution of A2, the following relationship between the VACF and MSD is employed,

$$\hat{W}(s) = \frac{2k_B T}{m} \frac{\hat{\psi}(s)}{s^2} \tag{A7}$$

where $\hat{W}(s)$ is the Laplace transform of MSD and $\hat{\psi}(s)$ is the Laplace transform of normalized VACF $\psi(t) = \frac{\langle v(t)v(0)\rangle_\tau}{\langle v^2\rangle_\tau}$. Therefore, the explicit form of MSD is,

$$\langle x^2\rangle = \frac{2k_BT}{m\omega_0^2}\left[1-\exp\left(-\frac{\gamma}{2}t\right)\left\{\cos(\hat{\omega}t)-\frac{\gamma}{2\hat{\omega}}\sin(\hat{\omega}t)\right\}\right] \tag{A8}$$

It is clear that as $t \gg 1/\gamma$,

$$\langle x^2\rangle_{t\gg 1/\gamma} = \frac{2k_BT}{m\omega_0^2} \tag{A9}$$

For a system with continuous oscillation driving sources, A2 becomes,

$$\dot{v}+\gamma v+\omega_0^2 x = r(t)+\sum_{Correlated} a_i\cos(\omega_i t) \tag{A10}$$

where the sum runs over all continuous oscillation frequencies. The solution of A10 is, derived using the same technique as the above on the homogeneous equation A4,

$$\langle x^2\rangle = \frac{2k_BT}{m\omega_0^2}\left[1-\exp\left(-\frac{\gamma}{2}t\right)\left\{\cos(\hat{\omega}t)-\frac{\gamma}{2\hat{\omega}}\sin(\hat{\omega}t)\right\}\right]+\sum_{Corr}C_i\cos(\omega_i t+\varphi_i) \tag{A11}$$

When the driving oscillation frequency is greater than the resonance frequency, $\omega_i \gg \omega_0$, and its amplitude is smaller $C_i \ll \frac{2k_BT}{m\omega_0^2}$, the oscillation driven MSD, A11, is just a small sinusoidal oscillation imposed on the MSD solution of A8. Therefore, as $t \gg 1/\gamma$, the average value of A11 is the same as in A9.